\documentclass[aps,floats,nofootinbib,amssymb,preprint,superscriptaddress]{revtex4}
\usepackage[sort&compress]{natbib}
\usepackage{epsf,epsfig}
\usepackage{amsmath}
\usepackage{hyperref}
\usepackage{subfigure}
\usepackage{graphicx}
\usepackage{color}
\usepackage{hyperref}

\newcommand{\be}{\begin{equation}}
\newcommand{\ee}{\end{equation}}
\newcommand{\bea}{\begin{eqnarray}}
\newcommand{\eea}{\end{eqnarray}}
\newcommand{\nn}{\nonumber}

\renewcommand{\check}{({\bf This statement needs to be checked!!})}

\newcommand{\ba}{\begin{array}}
\newcommand{\ea}{\end{array}}
\newcommand{\bi}{\begin{itemize}}
\newcommand{\ei}{\end{itemize}}
\newcommand{\ben}{\begin{enumerate}}
\newcommand{\een}{\end{enumerate}}

\newcommand{\cs}{\mathbb S}

\newcommand{\vsig}{$\left< v\sigma\right>$}
\newcommand{\vsigz}{$\left< v\sigma\right>_0$}
\newcommand{\degree}{^\circ}
\newcommand{\sci}[2]{#1$\times$10$^{\text{#2}}$}

\preprint{
\hbox to \hsize{
\hfill$\vcenter{\hbox{\bf MAD-PH-10-1561}
	\hbox{\bf NUHEP-TH/10-09}
	\hbox{\bf ANL-HEP-PR-10-39}}$}
}

\begin{document}
\title{\vspace*{.75in}
Light Higgs Boson, Light Dark Matter and Gamma Rays}
\author{Vernon Barger}
\affiliation{Department of Physics, University of Wisconsin, Madison, WI 53706}
\author{Yu Gao}
\affiliation{Department of Physics, University of Wisconsin, Madison, WI 53706}
\author{Mathew McCaskey}
\affiliation{Department of Physics, University of Wisconsin, Madison, WI 53706}
\author{Gabe Shaughnessy}
\affiliation{Northwestern University, Department of Physics and Astronomy, Evanston, IL 60208 USA}
\affiliation{HEP Division, Argonne National Lab, Argonne IL 60439 USA}

\thispagestyle{empty}

\begin{abstract}
A  light Higgs boson is preferred by  $M_W$ and $m_t$ measurements.  A complex scalar singlet addition to the Standard Model allows a better fit to these measurements through a new light singlet dominated state.  It then predicts a light Dark Matter (DM) particle that can explain the signals of DM scattering from nuclei in the CoGeNT and DAMA/LIBRA experiments.   Annihilations of this DM in the galactic halo, $AA\rightarrow b\bar{b}, c\bar{c}, \tau^+\tau^-$, lead to gamma rays that naturally improve a fit to the Fermi Large Area Telescope data in the central galactic regions. The associated light neutral Higgs boson may also be discovered at the Large Hadron Collider.

\end{abstract}
\date{\today}
\maketitle

\section{Introduction}
\label{sect:intro}

Our knowledge of the amount of Dark Matter (DM) in the Universe  has moved from a qualitative to a precision level from measurements of  Supernovae Ia, the Cosmic Microwave Background (CMB), radiation, Large Scale Structure (LLS)  and the Hubble constant.  The combined analysis of these datasets, including the WMAP 7-year CMB data, yields a mass density ratio, ${\Omega_M} = \rho_M/\rho_{critical}$, of $\Omega_{DM} h^2$ = 0.1109$\pm$0.0056 where $h$ = 0.71$\pm$0.025 is the Hubble constant in units of 100 km s$^{-1}$ Mpc$^{-1}$~\cite{Larson:2010gs}.  It is widely presumed that the DM is a stable, or nearly stable, elementary particle for which theoretical models provide many candidates.  Simulations of LLS tell us that the DM must be cold to seed large scale structure. Two principal categories  of cold DM particles are very light axions  and Weakly Interacting Massive Particles (WIMPs).  Searches for the axion of the Standard Model have set stringent limits.  

Much experimental effort is  being devoted to direct searches for WIMPs through their scattering from nuclei in underground detectors.  These DM particles are very non-relativistic.  The relative velocity of the WIMPs with respect to the nucleons is due to the motion of the Earth through the WIMP halo.   The underground experiments for WIMP detection record ionization, light, and/or phonons/heat from events.  The signals of nuclei recoiling from WIMP scattering can be distinguished from background events using the division of energy, timing and stopping power. A variety of detector materials have been employed, such as NaI, Ge, and Xe.  

Both Spin Independent (SI) and Spin-Dependent (SD) scattering have been studied and the present experimental reaches are about a factor of 10$^2$ lower for SI scattering.  From a theoretical vantage point, the experiments have reached interesting SI sensitivities.  

Already eight years ago, the DAMA collaboration, with a NaI detector at Gran Sasso, reported an annual modulation of event rates as evidence for the SI scattering of WIMPs. Newer DAMA/LIBRA data confirmed their earlier finding~\cite{Bernabei:2010mq}. The statistical significance of the combined DAMA/LIBRA data is 8.2 sigma. The DAMA/LIBRA signal corresponds a DM cross-section/nucleon of order $10^{-39}$ cm$^2$, at a WIMP mass of 5 GeV~\cite{Feng:2008dz,Feng:2008qn} with a signal band that extends roughly linearly down to \sci{5}{-42} cm$^2$ at a mass of 50 GeV~\cite{Bottino:2009km}. The band can be shifted by channeling, but it has been recently argued that such effects are small~\cite{Bozorgnia:2010xy}.  

The CoGeNT experiment, with a ultra-low noise Ge detector in the Soudan mine, reported a rising low energy spectrum that is unexplained by backgrounds.  This has been interpreted as a DM signal with a SI cross-section/nucleon just below  $10^{-40}$ cm$^2$ for $M_{DM}$ of 7 to 12 GeV~\cite{Andreas:2008xy,Chang:2010yk,Andreas:2010dz,Essig:2010ye,Graham:2010ca,Aalseth:2010vx}. 

In recoil experiments the quenching factors and other detection efficiencies in the relevant keV range are subject to systematic uncertainties, so  the boundary contours of the signal regions may be only approximate~\cite{Hooper:2010uy}.  For the efficiency assumptions of~\cite{Hooper:2010uy},
the inferred DAMA/LIBRA and CoGeNT regions meet at a DM mass of 7 GeV, for which the DM SI cross section is approximately \sci{2}{-40} cm$^2$.  

The null results found by the XENON10 and XENON100 experiments~\cite{Aprile:2010um} are compatible with the DM signal favored by the overlap of  DAMA/LIBRA and CoGeNT~\cite{Hooper:2010uy} after the uncertainties on the scintillation efficiencies of liquid Xenon are taken into account.~\cite{Sorensen:2010hq}.  The XENON data exclude the DAMA/LIBRA allowed region above a DM mass of 10 GeV.

Recently, the CRESST collaboration released preliminary data from their 400 kg-d run with nine 300g CaWO$_4$ crystal targets.  With a signal region defined by a recoil energy between 10-40 keV and a background dominated by $\alpha$ recoils, they estimate the total background to be $8.7\pm1.4$ events while they observe a total of 32 events~\cite{cresst-talk:2010}.  Such a signal event rate is consistent with a DM mass of $\lesssim 15$ GeV and a cross section ${\cal O}(10^{-41}\text{ pb})$~\cite{cresst-talk:2010}.

Taken together, these experimental results favor a DM candidate of mass near 10 GeV with a scattering cross section in the $\sigma_{SI}\sim {\cal O}(10^{-40}\text{ cm}^2)$ range.

The DM annihilations in the galaxy halo can be a source of energetic cosmic
rays. At a non-relativistic cross-section of \vsig$_0$ = 1 pb that gives
the right relic density, a light DM with mass of order 10 GeV can produce
gamma rays at a level that is detectable at the Fermi Gamma Ray Space Telescope
(FGST)~\cite{bib:fermi}. FGST in its scan mode measures the gamma ray energy
spectrum of the sky and a good agreement has been found with the expected astrophysics background. However, deviations
in the gamma ray spectrum expected from a power law background parameterization has
been seen in the FGST data near the galactic center~
\cite{bib:centralregion,Vitale:2009hr,Dobler:2009xz} and a DM contribution~\cite{Goodenough:2009gk,Cholis:2009gv} 
was shown to improve the agreement with the FGST data. Ref.~\cite{Goodenough:2009gk} shows that a 30 GeV DM particle provides the best fit to the FGST gamma ray data around the galactic center.

In the electroweak sector,  improvements in both the measurements and the SM calculations have reached the level of precision at which new physics contributions can be tested.  The CDF and D0 collaborations have provided the World's best measurements on the $W$-boson and $t$-quark masses which indirectly constrain new physics via loop corrections.  The loop contributions are sensitive to extensions of the Higgs sector.   In the SM there is a tension between the Higgs boson mass inferred from the electroweak precision observables (EWPO), which prefers a Higgs mass of ~90 GeV, and the lower LEP2 experimental bound of 114 GeV.  This tension can be alleviated by having a Higgs singlet that mixes with the SM Higgs doublet such that the their is a Higgs mass eigenstate that is below the LEP2 bound.

Thus, the DM signals point to a light DM particle and the EWPO measurements point to a light Higgs boson.  A minimal extension of the SM that can provide both of these particles is the Complex scalar singlet extension of the Standard Model (CSM).  

We previously showed that the SM with a complex singlet can be in good agreement with the CoGeNT and DAMA/LIBRA signals~\cite{Barger:2008jx}.  In the study, we show that it also provides a good fit to the precision observables $M_W$ and $m_t$ and to FGST gamma ray data in central galactic regions.

The remainder of this paper is organized as follows: In Section~\ref{sect:model}, we provide a brief overview of the complex scalar singlet model, while in Section~\ref{sect:ewpo}, we show that the scalar mass eigenstates provide a better fit to the observed $M_W$ and $m_t$ measurements by CDF and D0 than the SM can provide.  We discuss how this model can match the observed gamma ray excesses toward the center of the galaxy while maintaining the SI measurements in Section~\ref{sect:gamma}.  Finally, in Section~\ref{sect:summary}, we summarize and conclude.

\section{The Complex Scalar Singlet Model}
\label{sect:model}

A real scalar singlet~\cite{Silveira:1985rk,McDonald:1993ex,Burgess:2000yq,OConnell:2006wi,Barger:2007im,He:2009yd,He:2010nt}  added to the SM can either mix with the SM Higgs boson or be a Dark Matter particle.  The CSM allows both the mixing and a Dark Matter particle.  Assuming CP-conservation and including only renormalizable terms, the scalar potential of the CSM is~\cite{Barger:2008jx}
\begin{eqnarray}
V_\mathrm{csxSM}&=&\frac{m^{2}}{2}H^{\dagger}H+\frac{\lambda}{4}(H^{\dagger}H)^{2}+\frac{\delta_{2}}{2}H^{\dagger}H|\cs|^{2}+\frac{b_{2}}{2}|\cs|^{2}+\frac{d_{2}}{4}|\cs|^{4} \nonumber \\
&+&\left(\frac{-|b_{1}|}{4}  \cs^{2}+|a_1| \cs + c.c.\right),
\label{eq:potential}
\end{eqnarray}
where $H$ is the SM Higgs which obtains a vev $v=246$ GeV and $\cs = (S+iA)/\sqrt{2}$ is the complex singlet, with a vev $v_S$.  The $b_{1}$ term breaks a global $U(1)$ symmetric potential, giving mass to the DM $A$ state.  A non-zero $a_{1}$ avoids domain walls from an accidental $\cs\to-\cs$ symmetry. 

When the real component of the complex singlet obtains a vev, the $\delta_{2}$ term in the potential initiates mixing between $H$ and $S$.  The resulting coupling strengths of these Higgs eigenstates to the SM fermions and weak bosons are multiplied by the factors
\begin{equation}
g_{H_{1}}=
\cos\phi
\hspace{1cm} \text{and}\hspace{1cm}
g_{H_{2}}=-\sin\phi
,
\label{eq:twogs}
\end{equation}
resulting in a reduction in the production rate of the states, thereby allowing the lightest state to evade present SM bounds.  The mixing angle is given by the model parameters. 
The complex term in $V$ leads to a scalar field $A$ that is stable and is  thus the DM candidate.  The mass of the DM particle is determined by the parameters $b_{1}$ and $a_{1}$:
\begin{equation}
M_{A}^{2} = b_{1}-{\sqrt{2}a_{1}}/{v_{S}}.
\end{equation}
We use the scan and its associated constraints from Ref.~\cite{Barger:2008jx} as a guide through the parameter space and associated observables.


\section{Precision $M_W$ and $m_t$ measurements}
\label{sect:ewpo}

As noted above, global analyses of electroweak precision observables (EWPO)~\cite{Erler:2010wa} prefer a mass for the SM Higgs boson that is below the  direct lower bound of 114 GeV from LEP2 experiments.  Indeed, in the real singlet mixing case, an improvement to the oblique corrections can be found when the mass of the light state is reduced below the LEP SM Higgs limit~\cite{Barger:2007im}.  Moreover, a heavier $H_2$ state is allowed, up to $M_{H_2}=220$ GeV at the 95\% C.L. for maximal mixing.  However, as the $H_1$ state becomes dominantly singlet, this limit tightens to the SM limit of $M_{H_2}\approx 180$ GeV. As the CSM also provides a similar mixing scenario, an improved fit to the EWPO parameters is predicted in this model as well.  

Recently, the $W$-boson mass has been well measured by the CDF and D0 collaborations with about 1 fb$^{-1}$ of integrated luminosity to remarkable precision~\cite{Tevatron:2009nu}.  The world average, which includes a combination with the LEP II~\cite{Alcaraz:2006mx} result, is
\be
M_W^{meas} = 80.399\pm 0.023\text{ GeV},\\
\ee
while the top quark mass measurement by the CDF and D0 Collaborations with about 5.6 fb$^{-1}$ of integrated luminosity~\cite{TevatronElectroweakWorkingGroup:2010yx} is
\bea
m_t^{meas} &=& 173.3\pm1.1\text{GeV}.
\eea

The $W$-boson mass depends through radiative corrections on the top-quark mass, the $Z$-boson mass, the QED and QCD coupling constants, and the Higgs boson mass.  While these dependences are complicated, one can arrive at a reasonably accurate expansion in terms of the relevant parameters.  Using the expansion of the partial 3-loop calculation in Ref.~\cite{Awramik:2003rn} the dependences are
\be
\begin{array}{rcl}
M_W(M_H,m_t,M_Z, \Delta\alpha,\alpha_s) &=& M_W^0- c_1 dH-c_2 dH^2+c_3 dH^3+c_4(dh-1)\\
&-&c_5 d\alpha + c_6 dt-c_7 dt^2-c_8 dH dt + c_9 dh dt\\
&-&c_{10} d\alpha_s+c_{11}dZ,
\end{array}
\ee
where 
\bea
dH&=&ln\left({M_H \over \text{100 GeV}}\right),\quad dh = \left({M_H\over\text{100 GeV}}\right)^2,\quad dt=\left({m_t\over \text{174.3 GeV}}\right)^2-1,\nn\\
d\alpha&=&{\Delta\alpha\over 0.05907}-1,\quad d\alpha_s = {\alpha_s(M_Z)\over 0.119}-1,\quad dZ={M_Z\over \text{91.1875 GeV}} - 1.
\eea
The coefficients are given by
\bea
M_W^0 &=& 80.3799\text{ GeV},\quad c_1 = 0.05429\text{ GeV},\quad c_2 = 0.008939\text{ GeV},\quad c_3 = 0.0000890\text{ GeV},\nn\\
c_4 &=& 0.000161\text{ GeV},\quad c_5 = 1.070\text{ GeV},\quad c_6 = 0.5256\text{ GeV},\quad c_7 = 0.0678\text{ GeV},\\ 
c_8& =& 0.00179\text{ GeV},\quad  c_9 = 0.0000659\text{ GeV},\quad c_{10} = 0.0737\text{ GeV},\quad c_{11} = 114.9\text{ GeV},\nn
\eea
The current measured experimental values of the SM parameters are $\alpha_s(M_Z)=0.1184 \pm 0.0007$ and $M_Z =91.1876\pm 0.0021\text{ GeV}$~\cite{Amsler:2008zzb}.  The value $\Delta \alpha=\Delta \alpha_{lept}+\Delta \alpha_{had}$ is composed of separate hadronic and leptonic contributions, with $\Delta \alpha_{lept} = 0.031498$~\cite{Steinhauser:1998rq} and $\Delta \alpha_{had} = 0.02786\pm0.00012$~\cite{Amsler:2008zzb}.  This parameterization yields a value of $M_W$ that is accurate to 0.5 MeV for a SM Higgs boson masses up to 1 TeV~\cite{Awramik:2003rn}\footnote{The quality of the description of $M_W$ degrades gradually as the $H_2$ mass increases.}.

\begin{figure}[htbp]
\begin{center}
\includegraphics*[angle=0,width=0.49\textwidth]{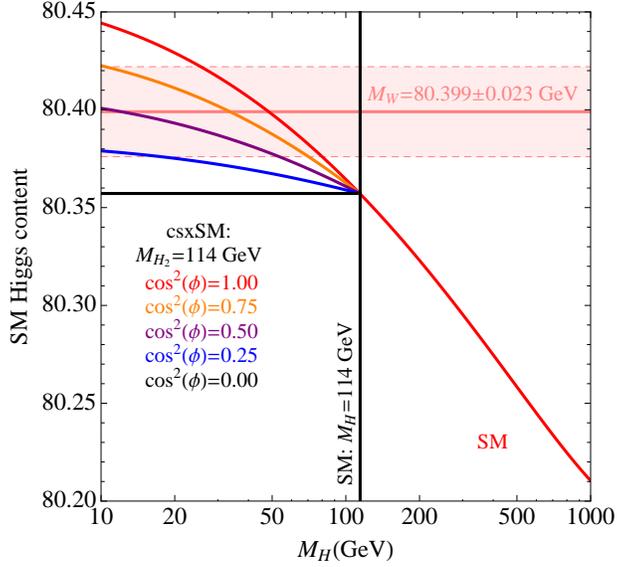}
\caption{The dependence of $M_W$ on $M_H$ for the SM (with $M_H$ above 114 GeV) and on $M_{H}\equiv M_{H_1}$ for the complex singlet model for various values of $\cos^2\phi$ with $M_{H_2}=114$ GeV.  The values of $m_t,M_Z, \Delta\alpha$ and $\alpha_s$ are fixed to their central values.}
\label{fig:locus}
\end{center}
\end{figure} 

The $W$-boson mass for the CSM with the SM content of Eq.~\ref{eq:twogs} is approximated by
\be
M_W(M_{H_1},M_{H_2},\phi,m_t) \approx \cos^2\phi M_W(M_{H_1},m_t)+\sin^2\phi M_W(M_{H_2},m_t),
\ee
where the additional dependences on $M_Z, \Delta\alpha$ and $\alpha_s$ are implicit~\footnote{This relation is approximate since the singlet contributions become non-trivial once the order of the calculation goes beyond 1-loop. However, since we use the precision measurements of $M_W$ and $m_t$ as a motivation for a light Higgs state, the precision of this calculation is not required to go to 3-loops.}. In Fig.~\ref{fig:locus}, we see the dependence of $M_W$ on $M_H$ for the SM (above $M_H=114$ GeV) and on $M_{H_1}$ for the CSM for varying values of $\cos^2\phi$ with $M_{H_2}=114$ GeV.  Generally, as the SM-content of the lightest Higgs increases for a given $H_1$ mass, the $W$-boson mass increases, allowing a better fit to the measured value. Moreover, for increasing $M_{H_2}$, the fits worsens.

Due to experimental constraints one cannot arbitrarily increase the SM-content of $H_1$. For a given light Higgs mass,  there is an upper bound from LEP2 on the amount of singlet-Higgs mixing through the measured limit of the $ZZh$ coupling~\cite{Sopczak:2001tg}.  As the Higgs mass decreases, its SM content must correspondingly decrease to suppress the production rate at LEP2, thereby mitigating the improvement of the $W$-boson mass prediction.

\begin{figure}[htbp]
\begin{center}
\includegraphics*[angle=0,width=0.49\textwidth]{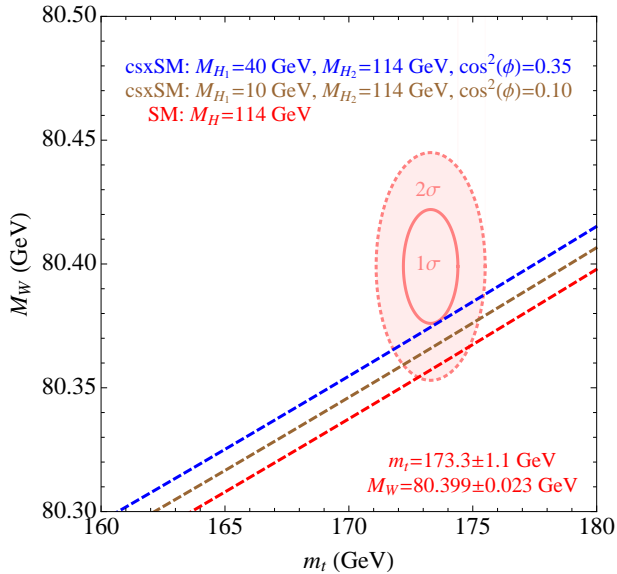}
\caption{Contours of the World's average  $m_t$ and $M_W$ at the $1\sigma$ and $2\sigma$ levels.  The SM (red dashed line) shows consistency with the measurements at about $2\sigma$.  The two CSM illustrations  (brown and blue dashed lines ) give an improved description of the data via the light Higgs state that has a singlet admixture characterized by the mixing angle $\phi$. }
\label{fig:mtmw}
\end{center}
\end{figure}

The contours in Fig.~\ref{fig:mtmw} represent the $M_W$ and $m_t$  measurements at one and two sigma.   Overlaid is the SM prediction which has an almost linear dependence in this narrow mass window.  Additionally, we show two CSM predictions with a light $H_1$ that is predominantly composed of a SM. These points  satisfy the various constraints from LEP2 experiments detailed in Ref.~\cite{Barger:2008jx}.  This demonstrates  that the CSM can describe the measured masses better than the SM by having a fraction of the Higgs contribution to the $W$-boson mass  come from the lighter singlet dominated state.  Nonetheless, a light $H_1$ state will fit the measured $W$-boson and top-quark masses better if it has a larger SM Higgs component, as  the $M_{H_1}=40$ GeV and $\text{cos}^2\phi=0.35$ case shows.

\begin{figure}[htbp]
\begin{center}
\includegraphics*[angle=0,width=0.49\textwidth]{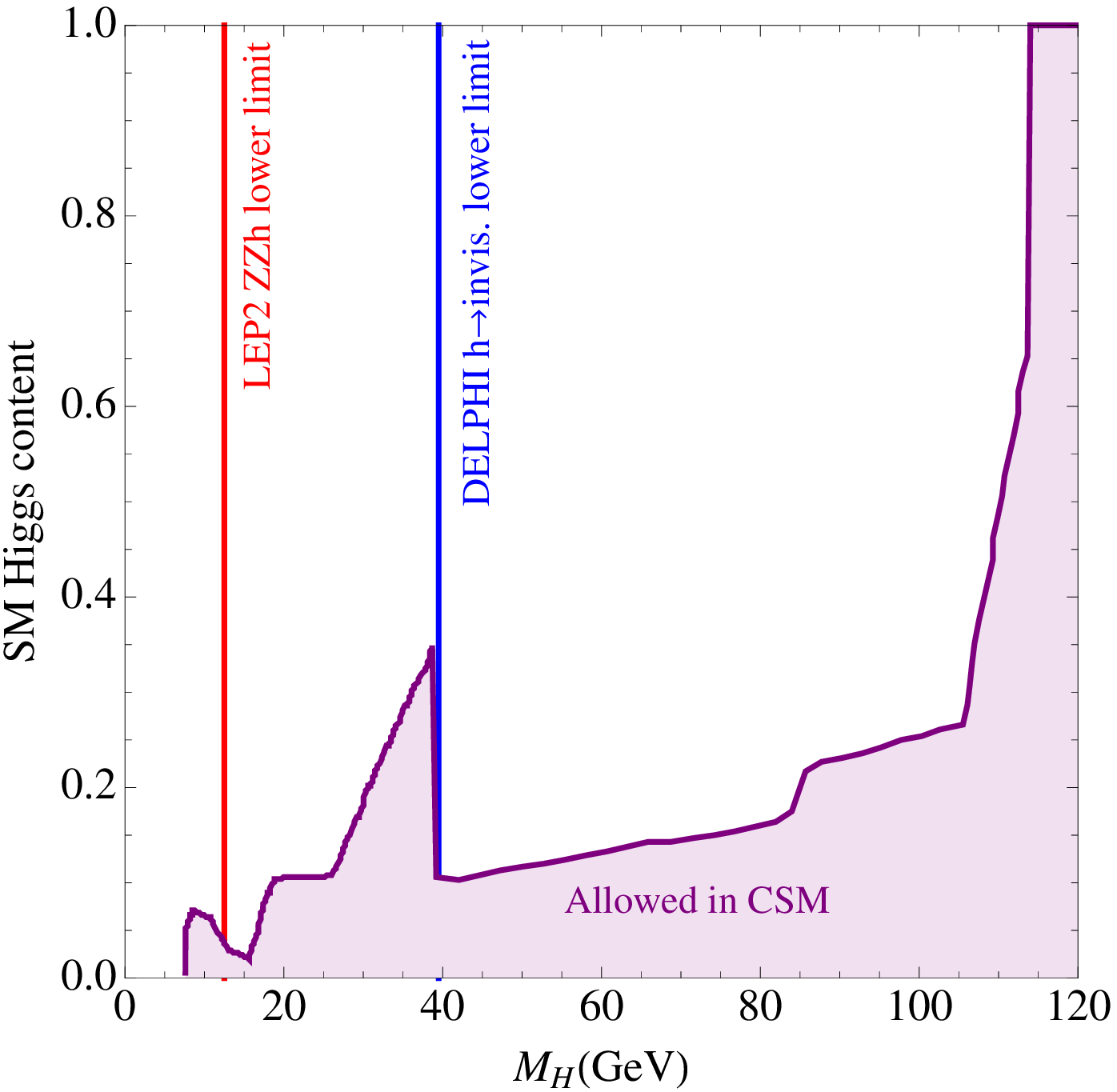}
\includegraphics*[angle=0,width=0.49\textwidth]{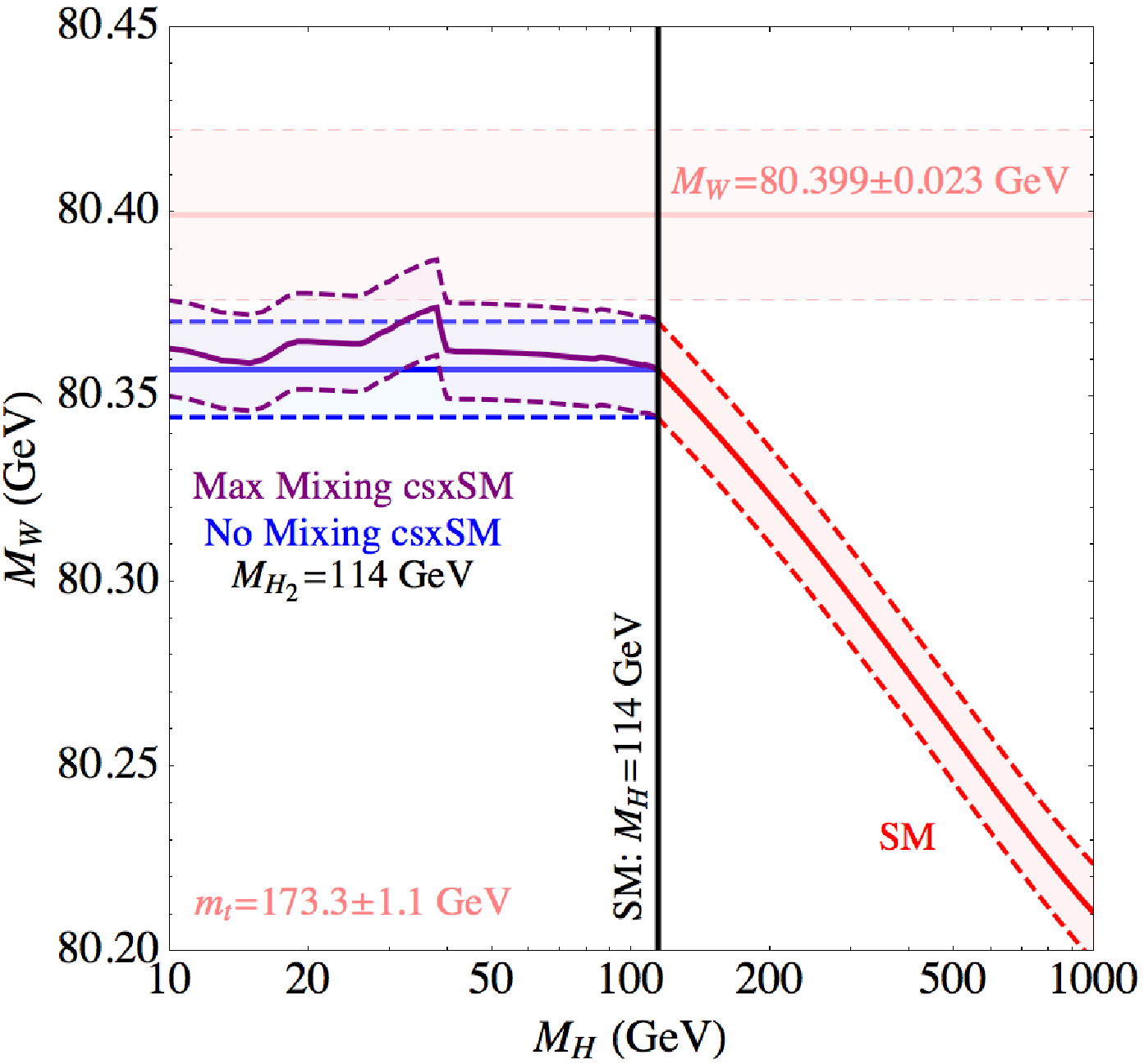}
\caption{The (a) maximal SM Higgs content of the lightest Higgs after LEP constraints are applied and (b) the predicted $W$-boson mass given in the maximal-mixing and no-mixing cases.}
\label{fig:minmax}
\end{center}
\end{figure}

The allowed values of the SM Higgs content upper bounding curve for $M_{H_1}<114$ GeV are shown by the shaded region in Fig.~\ref{fig:minmax}a.  These ranges of mixing satisfy various constraints from LEP2.  The $ZZh$ coupling is restricted below the SM expectation within the range 12-114 GeV, with the lower limit shown in Fig.~\ref{fig:minmax}a.  Limits on nonstandard Higgs boson decay modes that arise in singlet models are also included~\cite{Abbiendi:2002qp, Schael:2010aw}; these may be probed further at the Large Hadron Collier (LHC)~\cite{Barger:2006sk,Carena:2007jk,Cheung:2007sva,Chang:2008cw}.  The invisible decay is constrained by the combined LEP2 analysis in the 90-120 GeV mass range~\cite{delphi:2001xz}, while a more dedicated analysis from the DELPHI collaboration probes masses down to 40 GeV~\cite{Abdallah:2003ry}, below which the SM component of the $H_1$ state is allowed to increase somewhat, in turn giving a better fit to the $M_W$: see Fig.~\ref{fig:minmax}b \footnote{However, it is likely that an analysis of the LEP2 data in the mass range $M_H<$40 GeV would make the peak less prominent.}. Above $M_{H_1}=114$ GeV, the SM line is shown;  the band includes the dependence on the top-quark mass uncertainty.  Below 114 GeV, the dependence of $M_W$ on $M_{H_1}$ (assuming $M_{H_2}=114$ GeV) for zero mixing is given by the flat blue band, whereas the maximal mixing allowed by the LEP2 constraints is represented by the purple band.  Again, as the SM content of the lightest Higgs increases, the predicted $W$-boson mass is closer to the experimentally measured value denoted by the horizontal pink band.

Thus we have shown that the $W$-boson mass prediction within the content of the CSM, or more generally any model with additional singlet states that mix with the SM Higgs boson, can be in better agreement with the measured Tevatron $M_W$ and $m_t$ measurements and LEP2 constraints  than the SM.

\section{Gamma Ray \& Nucleon Recoil Signals}
\label{sect:gamma}

The Fermi Gamma Ray Space Telescope (FGST)~\cite{Abdo:2010nz} has measured energetic electron and gamma rays over large regions of the sky. Very good agreement of the data with galactic backgrounds is found except for gamma rays  above 1 GeV from angular regions close to the center of the Milky Way. Significant deviations are found in central galactic regions~\cite{bib:centralregion,Vitale:2009hr} from typical power law energy dependences expected from inverse Compton scattering and $\pi^0$ decays from astrophysical sources. A recent study has considered the annihilation of WIMPs with a 30 GeV mass as the source of excessive gamma ray in small angular regions ($<3\degree$) near the galactic center~\cite{Goodenough:2009gk}.

Annihilations of a light DM particle may explain the excess of diffuse GeV gamma rays near the galactic center (GC).  The $(\rho/M_{DM})^{2}$ enhancement of the DM source emissivity  with a cuspy DM profile can explain the DM signal enhancement near the GC.  A light DM particle can provide a cross section of the requisite size.

We consider annihilations of the light DM particle in the CSM as the origin of the observed gamma ray excess.  The differential gamma or electron flux is given by
\be 
\frac{d\phi^{\gamma,e}({r})}{dE}= \frac{\text{BF}}{2}\left(\frac{\rho({r})}{M_A}\right)^2 \left<v\sigma\right>
\sum_{i}\text{AF}_i\frac{d\phi^{\gamma,e}_i}{dE},
\label{eq:ann_rate}
\ee
where $i$ sums over the annihilation channels, \vsigz ~is the inclusive non-relativistic annihilation rate, AF$_i$ is the annihilation fraction into mode $i$ and $d\phi_i/dE$ denotes the differential photon and $e^+/e^-$ spectra of each annihilation channel. The dark matter distribution for which we adopt the cuspy Einasto~\cite{Navarro:2008kc} profile,
\be
\begin{array}{cc} 
\rho({r})=\rho_{\odot}\text{exp}\lbrace-\frac{2}{\alpha}[(r^{\alpha}-r_{\odot}^\alpha)/r_s^\alpha]\rbrace   &\hspace{1cm}\alpha=1.7,\ r_s=25 \text{kpc},\\
\end{array}
\ee
where the local dark matter density $\rho_{\odot}=0.3 \text{ GeV/cm}^3$.
As the CSM has pure $s$-channel  annihilation via the Higgs bosons, the dominant annihilation channels are the massive fermions: $AA\rightarrow b\bar{b}, \tau^+\tau^-, c\bar{c}$ with annihilation fractions that are proportional to the squares of their respective masses, $m_b^2:m^2_\tau:m_c^2$. 

The `boost factor' (BF) in Eq.~\ref{eq:ann_rate} normally refer to mechanisms that enhances the DM annihilation rate, such as Sommerfeld effect~\cite{Hisano:2003ec,Bovy:2009zs,Lattanzi:2008qa} of DM halos. However, in CSM the WIMP candidate $A$ has no excitation states or coupling to any light vector field, thus no major boost factor is expected to the non-relativistic annihilation cross-section, although there is some uncertainty associated with the choice of the DM halo distribution.

The total non-relativistic annihilation cross section is given by~\footnote{ Ignoring the loop-level $AA\rightarrow g g$ channel which is at the percent level.}
\be 
\begin{array}{rl} 
\left.\left<v\sigma \right>\right._{0}=&\frac{1}{4 \pi}\sum_{f} N^c_f\left(1-\frac{M^2_{f}}{M^2_A} \right)^{3/2}
\left(\frac{Y_{fH_1}g_{H_1AA}}{P_1}+\frac{Y_{fH_2}g_{H_2AA}}{P_2}\right)^2,
\end{array}
\label{eq:vsig}
\ee
where $f$ sums over heavy leptons and quarks; the color factor is $N^c_f=3$ for quarks and 1 for leptons;  $P_{1,2}$ are short-hand notations for the $s$-channel propagator and can be approximated as 
\be 
P_{1,2}= 4M_A^2-M^2_{H_{1,2}},
\label{eq:prop}
\ee
since $H_{1,2}$ are always off-resonance in the regions of interest.  The couplings in Eq.~\ref{eq:vsig} are parametrized by 
\be
\begin{array}{lcl} 
Y_{fH1}=-\cos{\phi}M_f/v, &\hspace{0.4cm}& Y_{fH2}=\sin{\phi}M_f/v, \\
g_{H_1AA}=(\delta_2 v \cos{\phi}+d_2 v_s \sin{\phi})/2,&&g_{H_2AA}=(\delta_2 v_s \cos{\phi}-d_2 v \sin{\phi})/2, 
\end{array}
\ee

To calculate the DM relic density, we used the MicrOmegas~\cite{Belanger:2010gh} package.  The SI scattering cross section for the CSM is~\cite{Barger:2010yn}
\be 
\sigma_{\text{SI}} = \frac{m_p^4 f^2_{tot}}{2\pi v^2 (m_p+M_A)^2}  \left(\frac{g_{H_1AA}g_{H_1}}{M_{H_1}^2}+
\frac{g_{H_2AA}g_{H_2}}{M_{H_2}^2}\right)^2,
\ee
where $m_p$ is the proton mass and $f_{tot}=0.350$ is the sum of integrated parton distribution for gluon and quarks inside protons~\cite{Ellis:2000ds}. There is no spin-dependent scattering in the CSM. 

\begin{figure}
\includegraphics[scale=0.7]{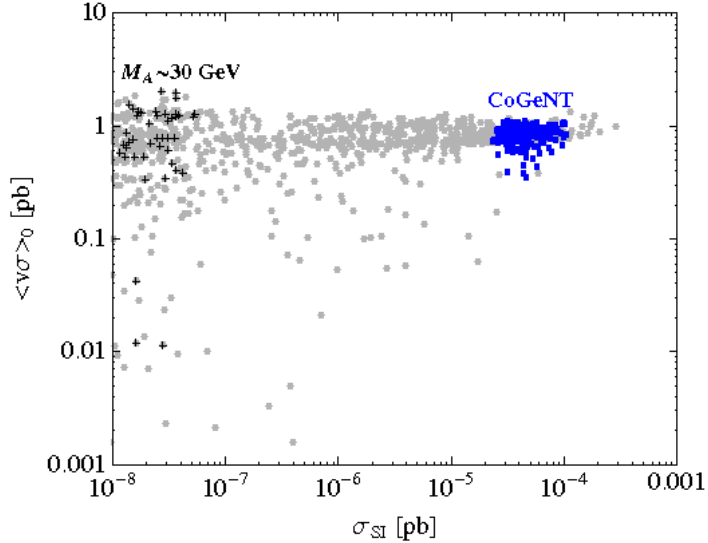}
\caption{
The CSM population (all points) on the \vsigz~ vs $\sigma_{SI}$ plane that reproduces the relic density observed by WMAP7 and the XENON100 exclusion limit on $\sigma_{SI}$. The CoGeNT allowed points are marked as blue dots (right).  The points with 27$<M_A<$33 GeV are marked a black `+'s (left) that better explains the FGST gamma ray spectra.
}
\label{fig:showrelic}
\end{figure} 
Fig.~\ref{fig:showrelic} illustrates the CSM population in the \vsigz ~versus $\sigma_{SI}$ plane that agree with relic density measurement from WMAP7 and XENON100 exclusion for $M_A$ less than 50 GeV.  The relic density constraint leads to a generic \vsigz~ near 1 pb that produces a gamma ray signal comparable to the galactic background.  At low $\sigma_{SI}$, $g_{HAA}$ is small and  \vsigz ~receives $s$-channel enhancement from $M_A$ being close to $M_{H_1}/2$.   In this scan we allowed the CSM parameters to vary over the following ranges 
\be  
\begin{array}{ccc}
5 \text{ GeV}< M_A <50 \text{ GeV},\ \ \ &
5 \text{ GeV}< \sqrt{b_1} < 500 \text{ GeV},\ \ \ &
10 \text{ GeV}<  v_s  < 1\text{ TeV},\\ 
0 < \lambda  < 2 ,&
-2 < \delta_2  < 2,&
0 <  d_2  <4.
\end{array}
\ee
Including CoGeNT and DAMA/LIBRA data,  the WIMP mass is further confined to a small range of $7-10$ GeV~\cite{Barger:2010yn}.

To show-case the gamma ray predictions, we choose $M_A=$ 10 GeV as typical of a CSM explanation of the CoGeNT and/or DAMA/LIBRA signals.  We also consider $M_A=$ 30 GeV, which has a SI cross section $\sigma_{SI}=$ \sci{1.4}{-44} cm$^2$ that is below the XENON100 bound, as an example of a somewhat higher $H_1$ mass.  The gamma signal consists of prompt photon emissivity given by Eq.~\ref{eq:ann_rate} and radiation associated with electrons produced in the DM annihilations, via their Inverse Compton scattering.

We test the dark matter contribution with a joint $\chi^2$ analysis of FGST $e^+e^-$ data~\cite{Abdo:2009zk} and gamma ray spectra in two large areas: (i) the `middle latitude' with
 $10\degree<|b|<20\degree$~\cite{Abdo:2010nz} and (ii) the `central' region with $|b|<5\degree$ and $|l|<30\degree$~\cite{bib:centralregion} 
 that includes the galactic center. In this analysis we do not take in account of possible correlations of the data from the two separate angular areas; this could have an effect on the systematics but should not alter the overall conclusions. The sizable angular coverage should smear out background fluctuations that may exist in very small regions.

\begin{table}[t]
\begin{tabular}{|ccccc|c|c|c|c|ccc|}
\hline
$\lambda$\ &\ $v_s$\ &\ $\delta_2$\ &\ $d_2$\ &\ $\sqrt{b_1}$\ &\ $M_A$\ &\ $M_{H_1}$\ &\vsigz ~(cm$^3$s$^{-1}$)&$\sigma_{SI}$(cm$^2$)&AF($b\bar{b}$)& AF($c\bar{c}$)& AF($\tau^+\tau^-$) \\
\hline
0.571&125&0.066&0.35&51&10&14&\sci{2.9}{-26}&\sci{3.1}{-41}&0.87&0.05&0.07\\
0.97&162&-0.16&0.38&51&30&55&\sci{4.6}{-26}&\sci{1.4}{-44}&0.86&0.04&0.07\\
\hline
\end{tabular}
\caption{Sample points for the CSM. The lower $M_A$ sits inside CoGeNT bound while the higher $M_A$ gives better agreement with Fermi gamma ray spectra. Both sample points are consistent with DM relic density and XENON100 $\sigma_{SI}$ constraints. The DM annihilation rate is largely determined by $A$ and $H_1$ masses.  
}
\label{tab:cases}
\end{table}

\begin{figure}[t]
\includegraphics[scale=0.6]{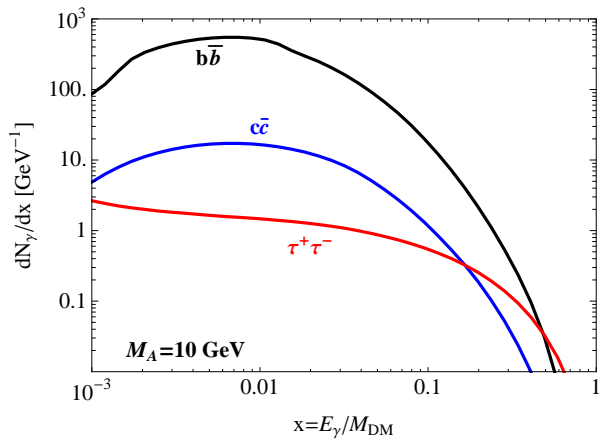}
\caption{Differential gamma ray spectra from individual channels for a 10 GeV $M_A$. Although sub-dominant  to the $b\bar{b}$ channel, the $\tau^+\tau^-$ channel produces more hard photons at energy fraction $x>0.5$ through copious $\pi^0$ decays}
\label{fig:comp}
\end{figure}

The galactic background and the dark matter induced gamma ray and electron signals are numerically evaluated with GALPROP package~\cite{Strong:1999sv,Strong:2001gh}. For the galactic diffuse background we assume power-law injection spectra $E^{-2.42}$ for nuclei and $E^{-\alpha_e}$ for the astrophysical electron background, where $\alpha_e$ was allowed to vary along with five other variables that parametrize the diffusion process, astrophysical electron background and the measured electron energy. See Ref.~\cite{Barger:2009yt} for the detailed numerical simulation.  The total DM annihilation rate was also treated as a free parameter.  
The annihilation channel branchings for two sample DM masses are listed in Table.~\ref{tab:cases}.   Fig.~\ref{fig:comp} illustrates the gamma ray contributions from the individual annihilation channels.  The gamma rays originate from neutral pions. The $b$ and $c$-quarks yield softer gamma rays than the $\tau^\pm$ leptons.

\begin{figure}[t]
\includegraphics[scale=0.85]{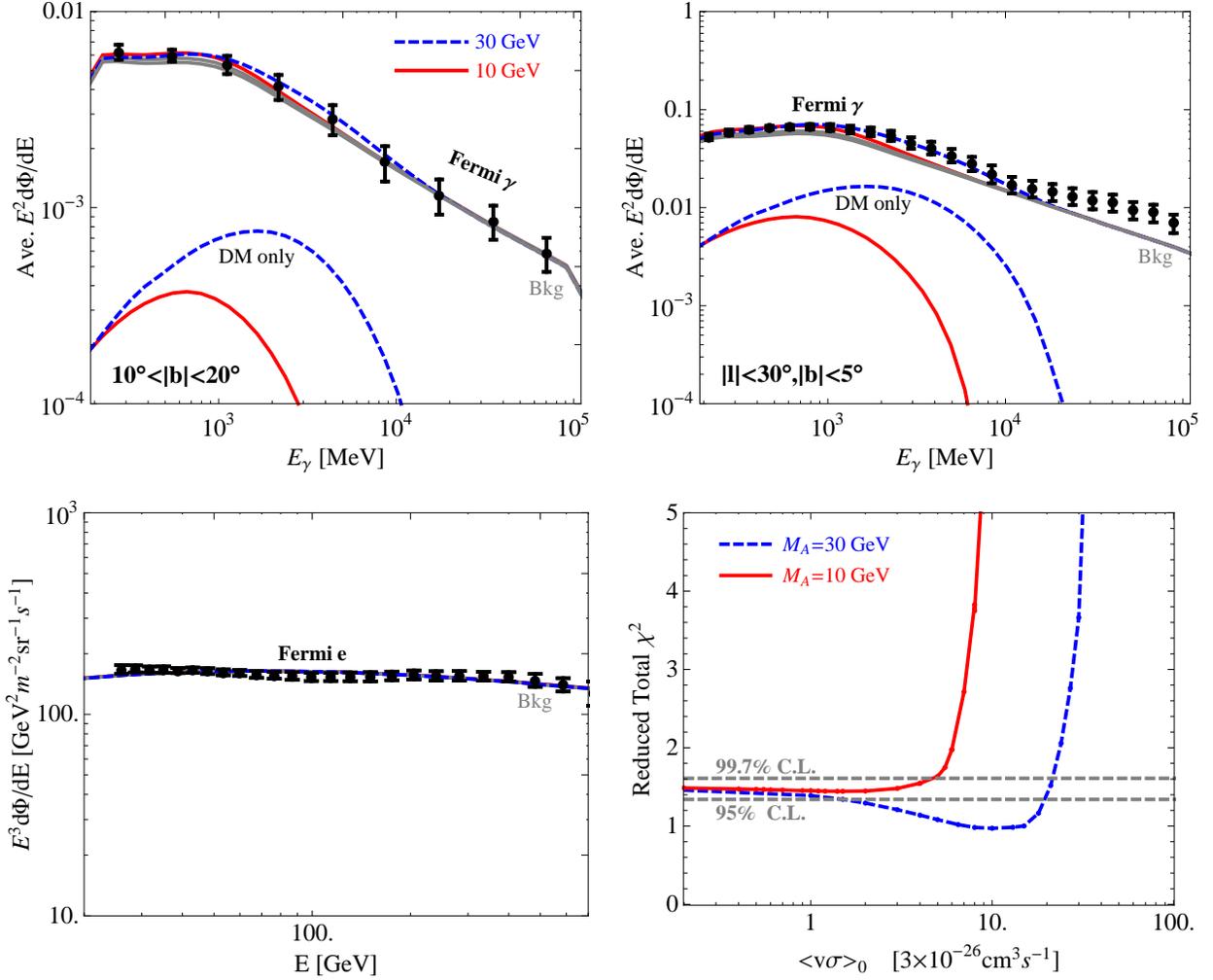}
\caption{Best fit scenarios for 10 GeV and 30 GeV DM masses. The angular regions are noted in the panels. The gamma ray contribution from dark matter enhances the total flux in the range of 1$\sim$10 GeV where an excess over a power-law galactic diffuse background has been observed. The Fermi electron data in the bottom panel are included in the fits. The right bottom panel shows the goodness of fit versus dark matter contribution rate for 10 GeV and 30 GeV DM masses. The number of degrees of freedom (DOF) is 54.}
\label{fig:gamma_fit}
\end{figure}

The WIMP mass is an important parameter in fitting the excessive gamma rays in the central galactic regions. $AA$ annihilation is dominated by the $b\bar{b}$ channel and the DM induced gamma ray $E_{\gamma}^2d\phi/dE_{\gamma}$ spectrum peaks at $E_{\gamma}=0.15 M_A$. Gamma rays from a $M_A$ below 10 GeV are likely to miss the energy range of the excess between 1 and 10 GeV. We find that the light dark matter at 30 GeV greatly improves the fit to Fermi gamma ray spectrum from a 2$\sigma$ background-only agreement to $\chi^2/dof$ less than 1. At 10 GeV or lower mass the DM induced gamma rays are relatively soft and lead to less effect in the fit to the FGST data.  Fig.~\ref{fig:gamma_fit} shows the minimal reduced $\chi^2$.

A good description of Fermi data gives
\be 
\left< v\sigma\right>_{Data} = 
\left\lbrace
\begin{array}{ll}
3\times 10^{-26}\left(\frac{10 \text{ GeV}}{M_A}\right)^2 \text{ cm}^3\text{s}^{-1},&\text{   for $M_A$=10 GeV }\\
3\times 10^{-25}\left(\frac{30 \text{ GeV}}{M_A}\right)^2 \text{ cm}^3\text{s}^{-1},&\text{   for $M_A$=30 GeV }\\
\end{array},
\right.
\label{eq:vsigdata}
\ee
and we can extrapolate the \vsigz ~over the narrow mass windows, using the predicted $M_A$ given by Eq.~\ref{eq:ann_rate}. At very light $M_A$, below $\sim10$ GeV, the gamma ray spectra from DM annihilations are too soft to explain the photon excess above 1 GeV, but the model still gives a fit with $\chi^2/dof< 2.0$ for \vsig$_0$ up to 6 pb. In this case, we choose \vsig$_0$ = 1 pb as a non-boosted annihilation cross-section which is detectable by FGST. For the 30 GeV case we choose the best-fit annihilation cross-section. There are numerous parameter combinations in the CSM that give  a \vsigz ~that explains the Fermi data with a natural boost factor $BF=1$.

A 30 GeV dark matter mass gives an improved fit to the excess in the gamma ray spectrum with a best-fit \vsigz=\sci{3}{-25}cm$^3$s$^{-1}$; this gives a boost factor above 3 that is necessary to reach the best fit gamma ray signal level. 

\begin{figure}[t]
\includegraphics[scale=0.75]{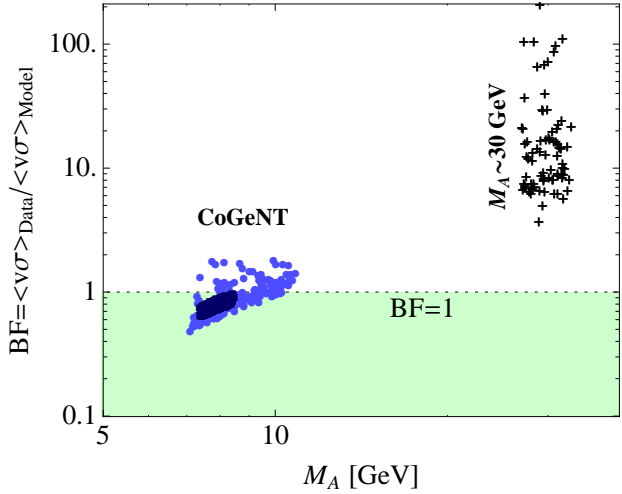}
\caption{ The ratio of the annihilation rate fit in Eq.~\ref{eq:vsigdata}, $\left< v\sigma\right>_{Data}$, to the  theoretical CSM model prediction, $\left< v\sigma\right>_{Model}$.  A ratio less than 1 means the model predicts a gamma ray signal larger than that allowed by the FGST data.   The dark and light blue dots represent separate CoGeNT allowed parameter regions that have a $H_1$ mass in the $7-12$ GeV and $30-60$ GeV ranges, respectively.  Each mass range has distinct collider signatures~\cite{Barger:2007im}. The CSM points with $M_A$ near 30 GeV (black `+'s) are also shown. Those points provide an improved fit to the FGST data, albeit with a boost factor of the order $10$.  All points satisfy the constraints in Fig.~\ref{fig:showrelic}.}
\label{fig:tot_xsec}
\end{figure}
Fig.~\ref{fig:tot_xsec} shows the non-relativistic \vsigz ~of $AA$ annihilation at the parameter  points that satisfy the XENON100 and relic density bounds. All the points in Fig.~\ref{fig:showrelic},~\ref{fig:tot_xsec} ~pass the $M_W$ and $m_t$ measurement constraints with less than $1.6 \sigma$.  We note that many points of the CSM parameter space naturally satisfy the constraints. Two regions of CMS parameter space~\cite{Barger:2007im} that are allowed by CoGeNT are shown as light and dark blue points. The light blue region satisfies $M_A<M_{H_1}<2M_A$ and a $M_A$ range of $7-12$ GeV while $M_{H_1}$ varies between 9 and 15 GeV . In the dark blue region $H_1$ has a broader range of $30 - 60$ GeV and is far off-shell,  $2 M_{A}< M_{H_1}$.  The DM mass in this region is also more concentrated near 8 GeV, allowing the invisible decay mode of the light Higgs $H_1\rightarrow AA$ to be open.

FGST will take data for 10 years.  The improved statistics will significantly improve its sensitivity to a DM annihilation source of gamma rays.  The contributions from a cross section of \vsig$_0$= 1 pb of a 10 GeV DM particle will produce a distinctive shape of the gamma ray energy spectrum
that will allow its confirmation or exclusion.

\section{Summary}
\label{sect:summary}

The Complex singlet extended Standard Model (CSM) has 3 scalar particles: two singlets and the neutral member of the SM Higgs doublet. The stable CP-odd singlet ($A$) is the dark matter and the CP-even singlet mixes with the SM Higgs boson. This model can provide a natural explanation of the possible CoGeNT and DAMA/LIBRA DM Spin Independent cross section signals with a DM mass in the $M_A\sim10$ GeV range.  There is then a light Higgs state ($H_1$) with a mass in either the $7-12$ GeV or the $30-60$ GeV range that is predominantly singlet, thereby allowing the light $H_1$ to escape detection at LEP2.  The heavier Higgs state ($H_2$) therefore has a predominantly SM doublet composition, with its couplings to SM particles universally reduced by a mixing factor and a mass range that is similar to that of the SM Higgs particle: $114-180$ GeV.  In this scenario, we have shown the new light Higgs boson allows a better fit to the precision observables $M_W$ and $m_t$ than the SM provides.  The $M_W$ mass is shifted to higher values than in the SM, in closer agreement with LEP2 and Tevatron measurements, improving from a nearly $2\sigma$ deviation in the SM to a $1\sigma$ deviation for $M_{H_2}=114$ GeV.  

The model predicts a DM annihilation contribution with \vsigz ~= 1 pb to gamma rays that explains the observed structure in the  1 - 10 GeV energy distribution of the Fermi diffuse gamma ray observations over the two large areas of the galaxy.  The gamma ray signals of the DM annihilations originate through the produced $b$ and $c$-quarks and $\tau^\pm$-leptons, through their subsequently decays to $\pi^0$s. The cuspy Einasto DM halo distribution yields the relative DM rates of the central galactic and mid-latitude data.  A even better fit to the high-latitude FGST data is obtained with $M_A=$ 30 GeV, albeit beyond the DM mass range allowed by XENON100.  Since the $H_1$ mass in the CSM is at most 60 GeV, the associated LHC Higgs boson signatures are potentially interesting: the $H_2$ can decay via SM modes, an invisible mode ($H_2\rightarrow AA$), and cascade modes, $H_2\rightarrow H_1H_1$.

\bigskip
{\bf Acknowledgments}

We thank D. Hooper for helpful information about Fermi data.
This work was supported in part by the U.S. Department of Energy Division of High Energy Physics under grants No. DE-FG02-95ER40896, DE-FG02-05ER41361, DE-FG02-08ER41531, DE-FG02-91ER40684 and Contract DE-AC02-06CH11357, by the Wisconsin Alumni Research Foundation, and by the National Science Foundation grant No. PHY-0503584.

\bibliography{csxsm-gamma}

\end{document}